\def\year{2021}\relax
\documentclass[letterpaper]{article} 
\usepackage{aaai21}  
\usepackage{times}  
\usepackage{helvet} 
\usepackage{courier}  
\usepackage[hyphens]{url}  
\usepackage{graphicx} 
\usepackage{natbib}  
\usepackage{caption} 

\usepackage{xcolor}

\usepackage{amsfonts}   
\usepackage{algorithmic}
\usepackage{algorithm}
\usepackage{multirow}
\usepackage{booktabs}

\usepackage{amsmath}

\usepackage{subfigure}

\title{MolGrow: A Graph Normalizing Flow for Hierarchical Molecular Generation}
\author {
        Maksim Kuznetsov,\textsuperscript{\rm 1}
        Daniil Polykovskiy, \textsuperscript{\rm 1} \\
}
\affiliations {
    \textsuperscript{\rm 1}  Insilico Medicine \\
    kuznetsov@insilico.com, daniil@insilico.com
}

\begin{document}
\maketitle

\begin{abstract}
We propose a hierarchical normalizing flow model for generating molecular graphs. The model produces new molecular structures from a single-node graph by recursively splitting every node into two. All operations are invertible and can be used as plug-and-play modules. The hierarchical nature of the latent codes allows for precise changes in the resulting graph: perturbations in the top layer cause global structural changes, while perturbations in the consequent layers change the resulting molecule marginally. The proposed model outperforms existing generative graph models on the distribution learning task. We also show successful experiments on global and constrained optimization of chemical properties using latent codes of the model.
\end{abstract}

\section{Introduction}
Drug discovery is a challenging multidisciplinary task that combines domain knowledge in chemistry, biology, and computational science. Recent works demonstrated successful applications of machine learning to the drug development process, including synthesis planning \cite{segler2018planning}, protein folding \cite{senior2020improved}, and hit discovery \cite{merk2018novo, zhavoronkov2019deep, kadurin2017cornucopia}. Advances in generative models enabled applications of machine learning to drug discovery, such as distribution learning and molecular property optimization. Distribution learning models train on a large dataset to produce novel compounds \cite{polykovskiy2018molecular}; property optimization models search the chemical space for molecules with desirable properties \cite{brown2019guacamol}. Often researchers combine these tasks: they first train a distribution learning model and then use its latent codes to optimize molecular properties \cite{Gomez-Bombarelli2018-jp}. For such models, proper latent codes are crucial for molecular space navigation.

We propose a new graph generative model---MolGrow. Starting with a single node, it iteratively splits every node into two. Our model is invertible and maps molecular structures onto a fixed-size hierarchical manifold. Top levels of the manifold define global structure, while the bottom levels influence local features.

Our contributions are three-fold:
\begin{itemize}
    \item We propose a hierarchical normalizing flow model for generating molecular graphs. The model gradually increases graph size during sampling, starting with a single node;
    \item We propose a fragment-oriented atom ordering that improves our model over commonly used breadth-first search ordering;
    \item We apply our model to distribution learning and property optimization tasks. We report distribution learning metrics (Fr\'echet ChemNet distance and fragment distribution) for graph generative models besides providing standard uniqueness and validity measures.
\end{itemize}

\begin{figure*}[t]
\begin{center}
    \centerline{\includegraphics[width=1.6\columnwidth]{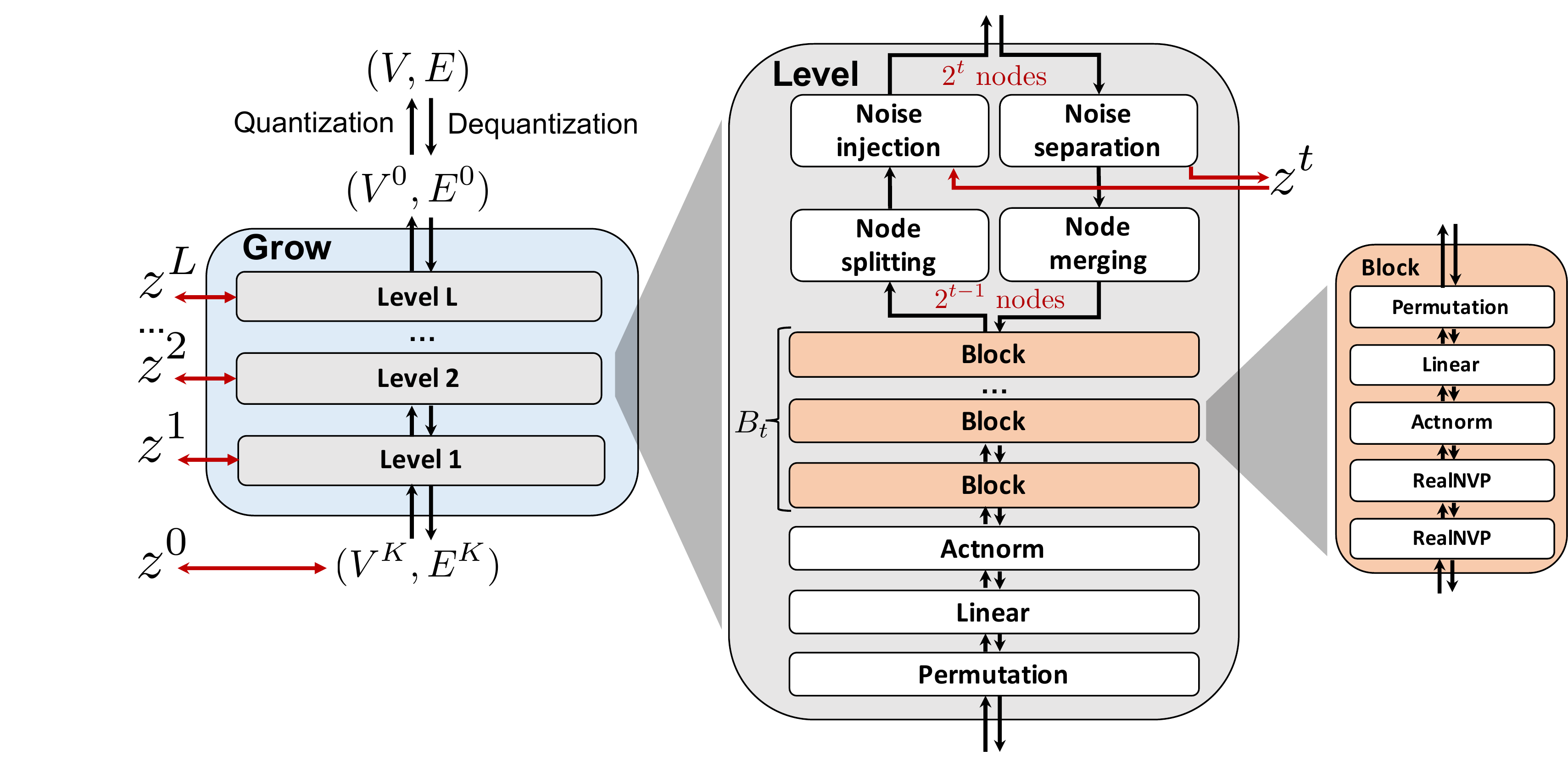}}
    \caption{MolGrow architecture. {\bf Left:} Full architecture combines multiple {\it levels} to generate latent codes $z^L, \dots, z^0$ from a graph $(V, E)$ and vice versa. {\bf Middle:} Each {\it level} separates noise, merges node pairs, applies multiple {\it blocks} and linear transformations; {\bf Right:} Each {\it block} applies three channel-wise transformations and two RealNVP layers.}
    \label{fig:grow}
\end{center}
\end{figure*}

\section{Background: Normalizing Flows} \label{sec:background}
Normalizing flows are generative models that transform a prior distribution $p(z)$ into a target distribution $p(x)$ by composing invertible functions $f_k$:
\begin{eqnarray}
    z = f_K \circ ... \circ f_{2} \circ f_1(x), \label{eq:forward} \\
    x = f^{-1}_1 \circ ... \circ f^{-1}_{K-1} \circ f^{-1}_K(z). \label{eq:inverse}
\end{eqnarray}
We call Equation~\ref{eq:forward} a forward path, and Equation~\ref{eq:inverse} an inverse path. The prior distribution $p(z)$ is often a standard multivariate normal distribution $\mathcal{N}(0, I)$. Such models are trained by maximizing training set log-likelihood using the change of variables formula:
\begin{equation}
    \log p(x) = \log p(z) + \sum\limits_{i=1}^{K} \log\left|\det \left(\frac{d h_i}{d h_{i-1}}\right)\right|,
    \label{eq:changeofvariable}
\end{equation}
where $h_i = f_i(h_{i-1})$, $h_0 = x$. To efficiently train the model and sample from it, inverse transformations and Jacobian determinants should be tractable and computationally efficient. In this work, we consider three types of layers: invertible linear layer, actnorm, and real-valued non-volume preserving transformation (RealNVP) \cite{dinh2016density}. We define these layers below for arbitrary $d$-dimensional vectors, and extend these layers for graph-structured data in the next section.

We consider an invertible linear layer parameterization by \citet{pmlr-v97-hoogeboom19a} that uses QR decomposition of a weight matrix: $h = QR \cdot z$, where $Q$ is an orthogonal matrix ($Q^T = Q^{-1}$), and $R$ is an upper triangular matrix with ones on the main diagonal. We use Householder reflections to parameterize Q:
\begin{equation}
    Q = \prod_{i=1}^{d'}\left(I - 2 \frac{v_iv_i^T}{\|v_i\|^2}\right),
\end{equation}
where $v_i$ are learnable column-vectors. The Jacobian determinant of a linear layer is $1$. There is an alternative way to formulate a linear layer with LU decomposition \cite{kingma2018glow}. However, in our experiments, QR decomposition showed more numerically stable results.

Actnorm layer \cite{kingma2018glow} is a linear layer with a diagonal weight matrix: $h = s \odot z + m$, where $\odot$ is an element-wise multiplication. Vectors $s$ and $m$ are initialized so that the output activations from this layer have zero mean and unit variance at the beginning of training. We use the first training batch for initialization. The Jacobian determinant of this layer is $\prod_{i=1}^d s_i$.

RealNVP layer \cite{dinh2016density} is a nonlinear invertible transformation. Consider a vector $z$ of length $d=2t$ with first half of the components denoted as $z_a$, and the second half as $z_b$. Then, RealNVP and its inverse transformations are:
\begin{gather}
 \begin{pmatrix} h_a \\ h_b \end{pmatrix}
 =
  \begin{pmatrix}
  z_b \\
   e^{s_{\theta}(z_b)} \odot z_a + t_{\theta}(z_b)
   \end{pmatrix} \\
   \begin{pmatrix} z_a \\ z_b \end{pmatrix}
 =
  \begin{pmatrix}
  \left(h_b - t_{\theta}(h_a)\right) / e^{s_{\theta}(h_a)} \\
  h_a
   \end{pmatrix}
\end{gather}
Functions $s_{\theta}$ and $t_{\theta}$ do not have to be invertible, and usually take form of a neural network. The Jacobian determinant of the RealNVP layer is $\prod_{i=1}^d e^{s_{\theta, i}(z_b)}$. We sequentially apply two RealNVP layers to transform both components of $z$. We also use permutation layer that deterministically shuffles input dimensions before RealNVP---this is equivalent to randomly splitting data into $a$ and $b$ parts.

\section{MolGrow (Molecular Graph Flow)}

In this section, we present our generative model---MolGrow. MolGrow is a hierarchical normalizing flow model (Figure~\ref{fig:grow}): it produces new molecular graphs from a single-node graph by recursively dividing every node into two. The final graph has $N = 2^{L}$ nodes, where $L$ is a number of node-splitting layers in the model. To generate graphs with fewer nodes, we add special padding atoms. We choose $N$ to be large enough to fit any graph from the training dataset.

We represent a graph with node attribute matrix $V \in \mathbb{R}^{N \times d_v}$ and edge attribute tensor $E \in \mathbb{R}^{N \times N \times d_e}$, where $d_v$ and $d_e$ are feature dimensions. For the input data, $V_i$ defines atom type and charge, $E_{i,j}$ defines edge type. Since molecular graphs are non-oriented, we preserve the symmetry constraint on all intermediate layers: $E_{i, j, k} = E_{j, i, k}$.

We illustrate MolGrow's architecture in Figure~\ref{fig:grow}. MolGrow consists of $L$ invertible {\it levels}, each {\it level} has its own latent code with a Gaussian prior. On a forward path, each {\it level} extracts the latent code and halves the graph size by merging node pairs. On the inverse path, each {\it level} does the opposite: it splits each node into two and adds additional noise. The final output on the forward path is a single-node graph graph $z^0 = (V^K, E^K)$ and latent codes from each {\it level}: $z^L, \dots, z^1$. We call $z^0$ a top level latent code.

\subsection{Dequantization}
To avoid fitting discrete graphs into a continuous density model, we dequantize the data using a uniform noise \cite{kingma2018glow}:
\begin{eqnarray}
    V^0_{i, j} = V_{i, j} + u^v_{i, j},\\
    E^0_{i, j, k} = E^0_{j, i, k} = \begin{cases}
    E_{i, j, k} + u^e_{i, j, k},  & i < j \\
    0, & i = j
    \end{cases}.
\end{eqnarray}
Elements of $u^v$ and $u^e$ are independent samples from a uniform distribution $\mathcal{U}[0, c]$. Such dequantization is invertible for $c \in [0, 1)$---original data can be reconstructed by rounding down the elements of $V^0_{i, j}$ and $E^0_{i, j, k}$. 
We dequantize the data for each training batch independently and train the model on $(V^0, E^0)$. Dequantizated graph $(V^0, E^0)$ is a complete graph.

\subsection{Node merging and splitting}

\begin{figure}[t]
\begin{center}
    \centerline{\includegraphics[width=0.8\columnwidth]{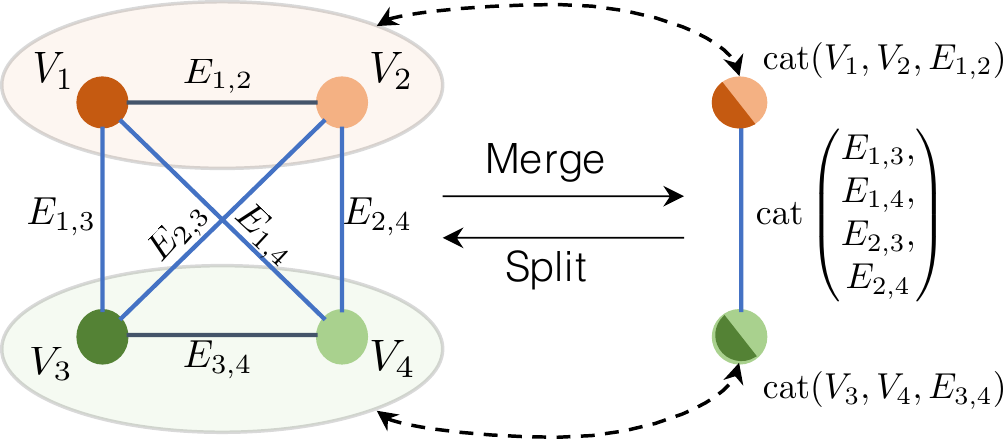}}
    \caption{Node merging and splitting example for a $4$-node graph. We concatenate features of nodes $V_1$ and $V_2$ and edge $E_{1,2}$ to get new node features. We also concatenate edge features $E_{1,3}$, $E_{1,4}$, $E_{2, 3}$, and $E_{2, 4}$. Splitting operation slices merged graph's node and edge features.}
    \label{fig:noise-divide}
\end{center}
\end{figure}

We use node merging and splitting operations to control the graph size. These operations are inverse of each other, and both operate by rearranging node and edge features. Consider a graph $(V^k, E^k)$ with $N_k$ nodes. Node merging operation joins nodes $2i$ and $2i+1$ into a single node by concatenating their features and features of the edge between them. We concatenate edge features connecting the merged nodes:
\begin{eqnarray}
    \underbrace{V^{k+1}_i}_{2d_{v}+d_{e}} =   \mathrm{cat}\big(
        \underbrace{V^{k}_{2i}}_{d_{v}},
        \underbrace{V^{k}_{2i+1}}_{d_{v}},
        \underbrace{E^{k}_{2i, 2i+1}}_{d_{e}}
    \big), \\
    \underbrace{E^{k+1}_{i, j}}_{4d_{e}} = \mathrm{cat}\big(
       \underbrace{E^{k}_{2i, 2j}}_{d_{e}},
       \underbrace{E^{k}_{2i, 2j+1}}_{d_{e}},
       \underbrace{E^{k}_{2i+1, 2j}}_{d_{e}},
       \underbrace{E^{k}_{2i+1, 2j+1}}_{d_{e}}
    \big).
\end{eqnarray}
Node splitting is the inverse of node merging layer: it slices features into original components. See an example in Figure~\ref{fig:noise-divide}.

\subsection{Noise separation and injection}
MolGrow produces a latent vector for each {\it level}. We derive the latent codes by separating half of the node and edge features before node merging and impose Gaussian prior on these latent codes. During generation, we sample the latent code from the prior and concatenate it with node and edge features. As we show in the experiments, latent codes on different levels affect the generated structure differently. Latent codes from smaller intermediate graphs (top level) influence global structure, while bottom {\it level} features define local structure.

\subsection{Block architecture}
The basic building block in MolGrow (denoted {\it block} in Figure~\ref{fig:grow}) consists of five layers. The first three layers (permutation, linear, and actnorm) serve as $1\times1$ convolutions. Each layer contains two transformations: one transforms every node and the other transforms every edge. The number of linear layer's Housholder reflections in matrix $Q$ is smaller than the dimension of $Q$. Hence, a combination of linear and permutation layers is not equivalent to a single linear layer.

The final two layers of the {\it block} are RealNVP layers.
RealNVP layer splits its input graph $(V^k, E^k)$  with $N_k$ nodes into $(V^{k, a}, E^{k, a})$ and $(V^{k, b}, E^{k, b})$ along features dimension. We transform $(V^{k, b}, E^{k, b})$ by projecting node and edge features onto a low-dimensional manifold and applying attention on complete graph edges (CAGE) architecture (Algorithm~\ref{alg:cage}). We compute the final output of RealNVP layer by applying fully-connected neural networks $s^v_{\theta}$, $t^v_{\theta}$, $s^e_{\theta}$, and $t^e_{\theta}$ to each node and edge independently:
\begin{eqnarray}
    (\overline{V}^{k, b}, \overline{E}^{k, b}) = \mathrm{CAGE}(V^{k, b}W_v, E^{k, b}W_e) \\
    V_i^{k+1, b} = \exp\left(s^{v}_{\theta}\left(\overline{V}_i^{k, b}\right)\right)\odot V_i^{k, a} + t^{v}_{\theta}\left(\overline{V}_i^{k, b}\right)\\
    V_i^{k+1, a} = V_i^{k, b} \\
    E_{i,j}^{k+1, b} = \exp\left(s^{e}_{\theta}\left(\overline{E}_{i,j}^{k, b}\right)\right) \odot E_{i,j}^{k, a} + t^{e}_{\theta}\left(\overline{E}_{i,j}^{k, b}\right)\\
    E_{i,j}^{k+1, a} = E_{i,j}^{k, b} 
\end{eqnarray}

Similar to other attentive graph convolutions \cite{velivckovic2017graph, guo-etal-2019-attention}, CAGE architecture uses a multi-head attention \cite{vaswani2017attention}. It also uses gated recurrent unit update function to stabilize training \cite{parisotto2019stabilizing}. Positional encoding in CAGE consists of two parts. First $d_v-\lceil\log_2N_k\rceil$ dimensions are standard sinusoidal positional encoding \cite{vaswani2017attention}:
\begin{eqnarray}
    \mathrm{pos}_{i, 2j} = \sin\left(i / 10000^{2j/d_v} \right), \\ 
    \mathrm{pos}_{i, 2j+1} = \cos\left(i / 10000^{2j/d_v}\right).
\end{eqnarray}
The last $\lceil\log_2N_k\rceil$ components of $\mathrm{pos}_i$ contain a binary code of $i$. We add multiple blocks before the first and after the last {\it level} in the full architecture.

\begin{algorithm}[h]
   \caption{Attention on complete graph edges (CAGE)}
   \label{alg:cage}
\begin{algorithmic}[1]
   \STATE {\bfseries Input:} Complete graph $(V, E)$ with node feature matrix $V \in \mathbb{R}^{n \times d_v}$ and edge feature tensor $E \in \mathbb{R}^{n \times n \times d_e}$.
  \STATE {\bfseries Output:} Transformed complete graph $(\overline{V}, \overline{E})$ of the same dimension as $(V, E)$.
  \STATE Compute positional encodings matrix $\mathrm{pos} \in \mathbb{R}^{n \times d_v}$
  \FOR{$i = 1$ {\bfseries to} $n$}
    \STATE Allocate $\mu \in \mathbb{R}^{n \times d_v}$
    \FOR{$j = 1$ {\bfseries to} $n$}
    \STATE $\mu_{j} = f_{\theta}\left(\mathrm{cat}\left(E_{i,j}, V_i, V_j\right)\right) + \mathrm{pos}_j$ --- compute message $j \to i$ of dimension $d_v$ using a fully-connected neural network $f_{\theta}$
    \ENDFOR
    \STATE $q = V_i + \mathrm{pos}_i$ --- compute attention query
    \STATE $r = \text{Multi-Head Attention}(\textrm{query}{=}q, \textrm{keys}{=}\mu, \textrm{values}{=}\mu)$ --- aggregate messages
    \STATE $\overline{V}_i = \textrm{GRU}_{V}(x{=}r,\, h{=}V_i)$ --- update node feature matrix using a GRU cell
    \STATE $\nu_{i, j} = \mathrm{cat}\left(E_{i,j},  \overline{V}_i, \overline{V}_j\right),~\forall j$ --- compute edge update vectors
    \STATE $\overline{E}_{i,j} = \frac12\textrm{GRU}_{E}(x{=}\nu_{i,j},\, h{=}E_{i,j}) + \frac12\textrm{GRU}_{E}(x{=}\nu_{j, i}, h{=}E_{j, i})~\forall j$ --- update edge features
 \ENDFOR
\end{algorithmic}
\end{algorithm}

\subsection{Layout and padding}

\begin{figure*}[ht]
\begin{center}
    \centerline{\includegraphics[width=1.4\columnwidth]{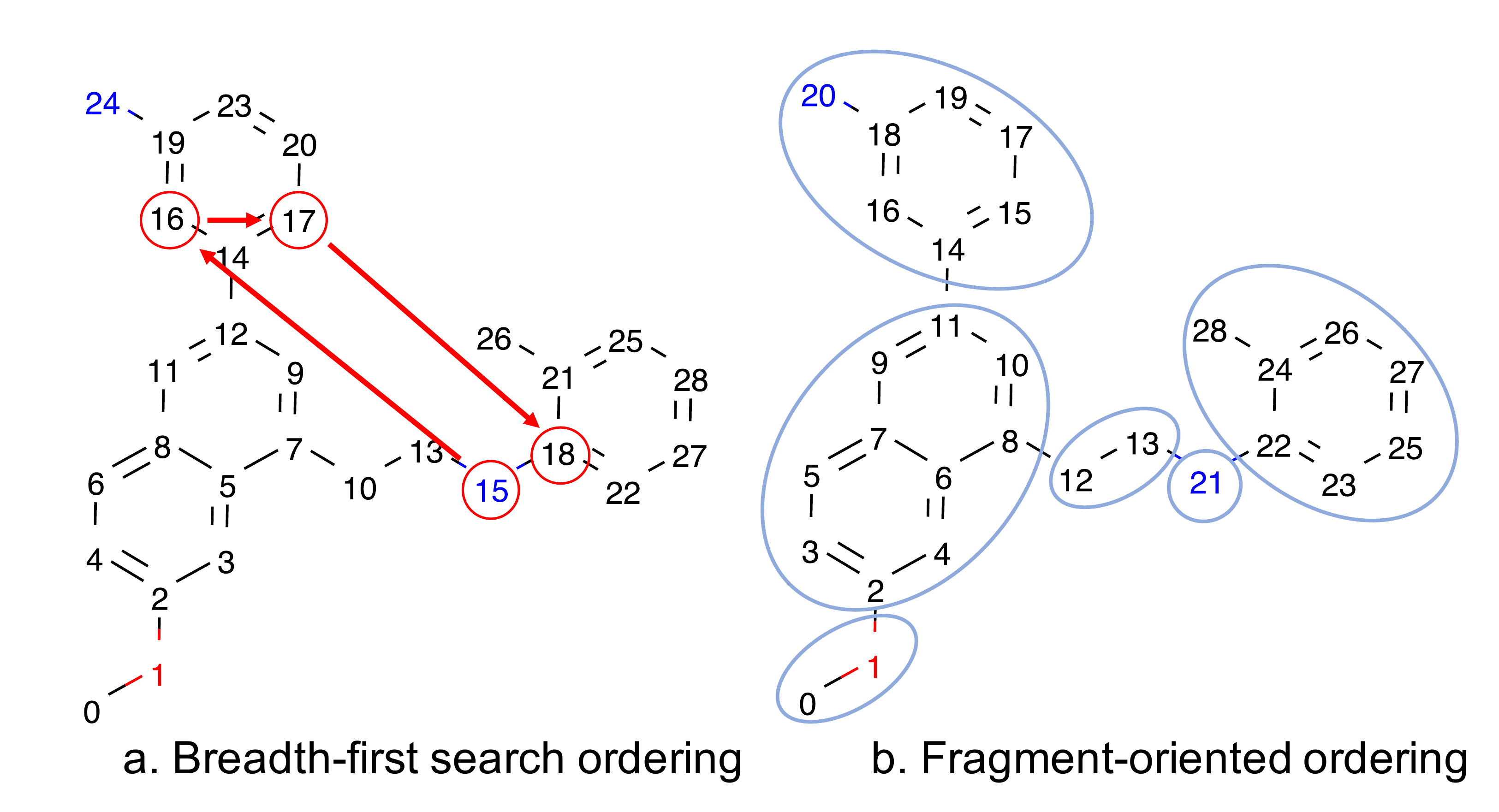}}
    \caption{Different atom orderings. Numbers are atom's indices in a particular ordering. Note that BFS ordering {\bf (a)} generates two fragments in parallel (see nodes $15$-$18$), while our method completes a fragment before transitioning to the next one. For fragment-oriented ordering {\bf (b)}, we circled extracted fragments.}
    \label{fig:layout}
\end{center}
\end{figure*}

Similar to the previous works \cite{shi2020graphaf, pmlr-v80-you18a}, we achieved better results when learning on a fixed atom ordering, instead of learning a distribution over all permutations. Previous works used breadth-first search (BFS) atom ordering, since it avoids long-range dependencies. However, BFS does not incorporate the knowledge of common fragments and can mix their atoms (Figure~\ref{fig:layout}a). We propose a new atom ordering to incorporate prior knowledge about frequent fragments. Our ordering better organizes the latent space and simplifies generation.

We break the molecule into fragments by removing BRICS \citep{Degen2008-fk} bonds and bonds connecting rings, linkers, and decorations in the Bemis-Murcko scaffold \cite{Bemis1996-js}. We then enumerate the fragments and atoms in each fragment using BFS ordering (Figure~\ref{fig:layout}b). We recursively choose padding positions, minimizing the number of edges after node merging layers (Algorithm~\ref{alg:padding}).

\begin{algorithm}[h]
   \caption{Balanced padding (function ``pad'')}
   \label{alg:padding}
\begin{algorithmic}[1]
   \STATE {\bfseries Input:} List of fragments $f_{1:K} = [f_1 \dots f_K]$, where $f_i$ contains atom indices in the $i$-th fragment; $N$---target graph size, power of $2$.
  \STATE {\bfseries Output:} Atom order with $*$ indicating padding positions ($*_N$ indicates $N$ sequential paddings).
  \IF{$K=1$ and $|f_1| \le N/2$}
  \STATE Randomly add padding: with 50\% probability {\bf return} $\mathrm{cat}(\mathrm{pad}(f_{1}, N/2), *_{N/2})$, otherwise {\bf return} $\mathrm{cat}(*_{N/2}, \mathrm{pad}(f_{1}, N/2))$
  \ENDIF
  \STATE Find possible splitting positions $B$: $b \in B$ if left and right parts fit into subtrees: $|f_{1:b}| \le N/2$ and $|f_{b+1:K}| \le N/2$ 
  \IF{$|B| > 0$}
  \STATE Sample any index $b \in B$ that minimizes the number of bonds between $f_{1:b}$ and $f_{b+1:K}$.
  \STATE Recursively add padding to left and right parts:
  \STATE {\bf return} $\mathrm{cat}\Big(\mathrm{pad}(f_{1:b}, N/2), \mathrm{pad}(f_{b+1:K}, N/2)\Big)$  \ELSE
  \STATE Add padding to the right
  \STATE {\bf return} $\mathrm{cat}(f_{1:K}, *_{N-|f_{1:K}|})$
  \ENDIF
\end{algorithmic}
\end{algorithm}

\begin{table*}[ht]
\caption{Distribution learning metrics on MOSES dataset.
}
\label{tab:moses}
\centering
\resizebox{2\columnwidth}{!}{
\begin{tabular}{lllllll}
\toprule
Method  & FCD/Test $(\downarrow)$ & Frag/Test $(\uparrow)$ & Unique@10k $(\uparrow)$ & Novelty $(\uparrow)$ 
\\
\midrule
\multicolumn{5}{c}{Graph-based models} \\ 
\midrule
MolecularRNN \cite{popova2019molecularrnn}  & 23.13 & 0.56   & 98.6\%       & 99.9\%
\\
GraphVAE \cite{graphVAE}    & 49.39 & 0.0         & 5\%        & {\bf 100\%}
\\
GraphNVP \cite{madhawa2019graphnvp}     & 29.95    & 0.62           & 99.7 \%          &  99.9\%
\\
GraphAF (BFS) \cite{shi2020graphaf}    & 21.84 & 0.651         & 97\%       & 99.9\%
\\
MoFlow \cite{zang2020moflow}    & 28.05 &  0.685        & 100\%       & 99.99\%
\\
\midrule
\multicolumn{5}{c}{Proposed model} \\
\midrule
MolGrow (fragment-oriented) & $\mathbf{ 6.284 \pm 0.986}$  & $0.929 \pm  0.025$  &  $99.28 \pm 0.62$\%      & $99.26 \pm 0.12$\%
\\
MolGrow (BFS)  & $9.962 \pm 0.795$ & $0.932 \pm 0.01$ & $\mathbf{100 \pm 0.0}$\% & $99.37 \pm 0.08$\%
\\
MolGrow (BFS on fragments)  & $16.15 \pm 1.026$ & $0.868 \pm 0.018$ & $\mathbf{100 \pm 0.0}$\% & $\mathbf{100 \pm 0.0}$\%
\\
MolGrow (random permutation) & $40.17 \pm 4.709$ & $0.051 \pm 0.034$ & $58.96 \pm 38.11$\% & $\mathbf{100 \pm 0.0}$\%
\\
MolGrow (GAT instead of CAGE) &  $6.523 \pm 0.302$  & $\mathbf{0.941 \pm 0.013}$    &  $99.36 \pm 0.3$\%      & $99.32 \pm 0.05$\%
\\
MolGrow (No positional embedding)&  $6.771 \pm 0.555$  & $0.937 \pm 0.006$   &  $99.49 \pm 0.19$\%      & $99.41 \pm 0.06$\%
\\
\midrule
\multicolumn{5}{c}{SMILES and fragment-based models} \\
\midrule
CharRNN (from MOSES benchmark) &  \textbf{0.073}  & \textbf{0.9998}  &  99.73\%      & 84.19\%
\\
VAE (from MOSES benchmark) &  0.099  & 0.9994  &  99.84\%      & 69.49\%
\\
JTN-VAE (from MOSES benchmark) &  0.422  & 0.9962  &  \textbf{100\%}      & \textbf{91.53\%}
\\
\bottomrule
\end{tabular}
}
\end{table*}

\begin{figure*}[ht]
\begin{center}
    \centerline{\includegraphics[width=0.95\textwidth]{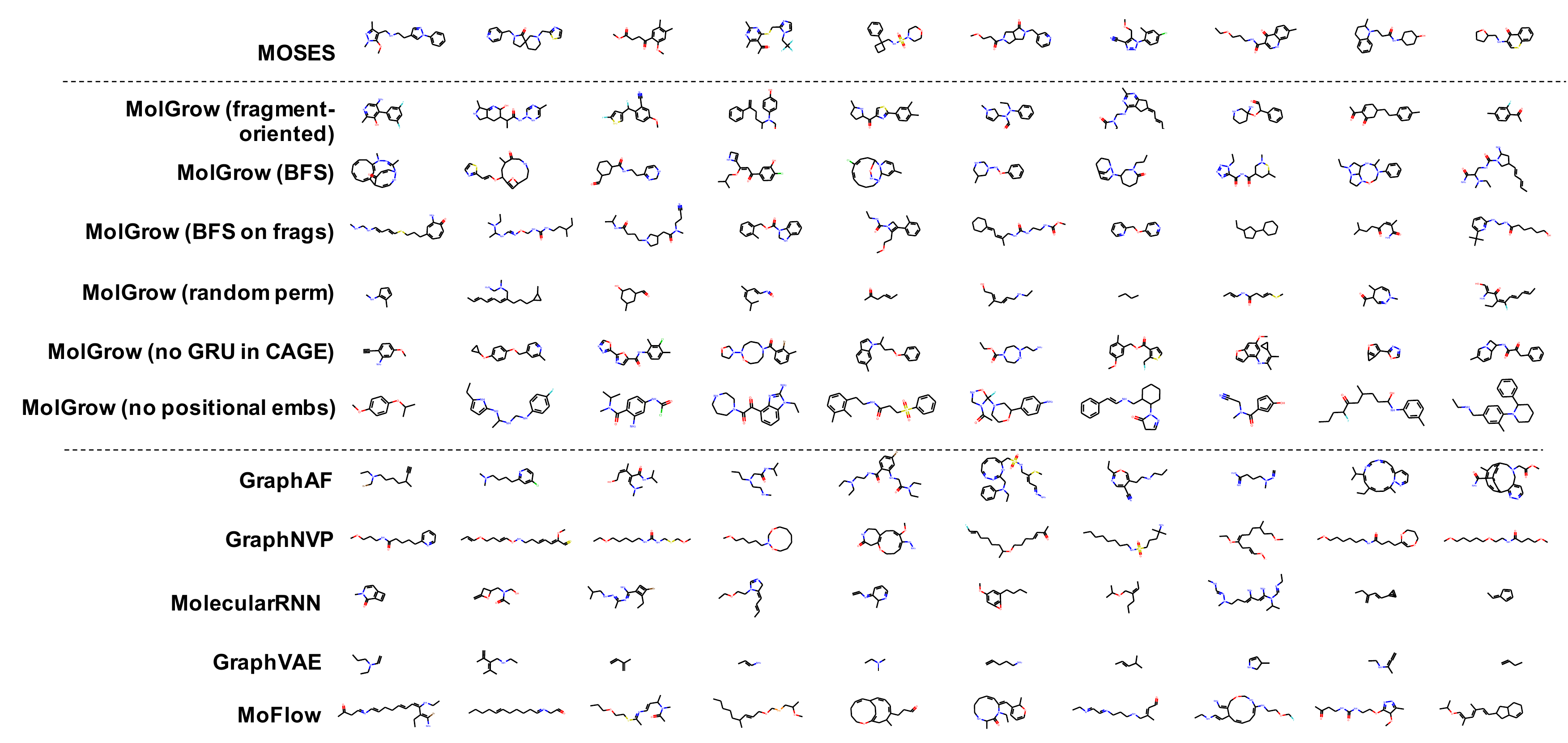}}
    \caption{Samples from molecular graph generative models trained on MOSES.}
    \label{fig:some_samples}
\end{center}
\end{figure*}

\section{Related work}
Many well-known generative models work out of the box for molecular generation task. By representing molecules as strings, one can apply any sequence generation model: language models \cite{Segler2018-eo}, variational autoencoders \cite{Gomez-Bombarelli2018-jp}, and generative adversarial networks \cite{Sanchez-Lengeling2017-sp}. Molecular graphs satisfy a formal set of rules: all atoms must have a proper valency, and a graph must have only one component.
These constraints can be learned implicitly from the data or explicitly by specifying grammar rules \cite{kusner2017grammar, OBoyle2018-fj, krenn2019selfies}.

Multiple generative models for molecular graphs were proposed. Graph recurrent neural network (GraphRNN) \cite{pmlr-v80-you18a} and molecular recurrent neural network (MolecularRNN) \cite{popova2019molecularrnn} use node and edge generators: node generator sequentially produces nodes; edge generator sequentially predicts edge types for all the previous nodes from the hidden states of a node generator. Molecular generative adversarial network (MolGAN) \cite{De_Cao2018-ju} trains a critic on generated graphs and passes the gradient to the generator using deep deterministic policy gradient \cite{lillicrap2015continuous}. Graph variational autoencoder (GraphVAE) \cite{graphVAE} encodes and decodes molecules using edge-conditioned graph convolutions \cite{simonovsky2017dynamic}. Graph autoregressive flow (GraphAF) \cite{shi2020graphaf} iteratively produces nodes and edges; discrete one-hot vectors are dequantized, and tokens are decoded using argmax. The most similar work to ours is a graph non-volume preserving transformation (GraphNVP) \cite{madhawa2019graphnvp} model. GraphNVP generation is not autoregressive: the model produces a dequantized adjacency matrix using normalizing flow, turns it into a discrete set of edges by computing argmax, and obtains atom types using a normalizing flow. MoFlow \cite{zang2020moflow} exploits a two-stage graph generation process similar to GraphNVP: first, it generates an adjacency matrix with Glow architecture and then recovers node attributes with additional coupling layers. Unlike GraphNVP and MoFlow, we generate the graph hierarchically and inject noise on multiple levels. We also produce nodes and edges simultaneously. \citet{tran2019discrete} model discrete data directly with straight-through gradient estimators. Application of such model to graph structured data is yet to be explored. 

All graph generative models mentioned above are not permutation invariant, and most of these models employ a fixed atom order. GraphVAE aligns generated and target graphs using Hungarian algorithm, but has high computational complexity. \citet{vinyals2015order} report that order matters in set to set transformation problems with specific orderings giving better results. Most models learn on a breadth-first search (BFS) atom ordering \cite{pmlr-v80-you18a}. In BFS, new nodes are connected only to the nodes produced on the current or previous BFS layers, avoiding long range dependencies. Alternatively, graph generative models could use a canonical depth-first search (DFS) order \cite{Weininger1989}. Note that unlike graph-based generators, string-based generators seem to improve from augmenting atom orders \cite{bjerrum2017smiles}.

Several works studied fragment-based molecular generation. \citet{Jin2018-sm} replace molecular fragments with nodes to form a junction tree. They then produce molecules by sampling a junction tree and then expanding it into a full molecule. The authors expanded this approach to incorporate larger fragments as tokens (motifs) \cite{jin2020hierarchical}.

\section{Experiments}

We consider three problems: distribution learning, global molecular property optimization and constrained optimization. For all the experiments, we provide model and optimization hyperparameters in supplementary material A; the source code for reproducing all the experiments is provided in supplementary materials. We consider hydrogen-depleted graphs, since hydrogens can be deduced from atom valence.

\subsection{Distribution learning}
In distribution learning task, we assess how well models capture the data distribution. We compare generated and a test sets using Fr\'echet ChemNet distance (FCD/Test) \cite{Preuer2018-pf}. FCD/Test is a Wasserstein-1 distance between Gaussian approximations of ChemNet's penultimate layer activations. We also computed cosine similarity between fragment frequency vectors in the generated and test sets. We report the results on MOSES \cite{polykovskiy2018molecular} dataset in Table~\ref{tab:moses}. MolGrow outperforms previous node-level graph generators by a large margin. Note that SMILES-based generators (CharRNN and VAE) and fragment-level generator (JTN-VAE) outperform all node-level graph models. We hypothesize that such representations impose strong prior on the generated structures. We provide samples from graph-based models in Figure~\ref{fig:some_samples}. Note that baseline models tend to produce macrocycles which were not present in the training set; molecules produced with GraphNVP contain too few rings. Ablation study demonstrates the advantage of fragment-oriented ordering and CAGE over standard graph attention network GAT \cite{velivckovic2017graph}.
We provide the results on QM9 \cite{ramakrishnan2014quantum} and ZINC250k \cite{kusner2017grammar} datasets in supplementary material B. In Figure 6 of Supplementary materials, we show how resampling different latent codes affect the generated structure.

\subsection{Global optimization}
The goal of a global optimization task is to produce new molecules that maximize a given chemical property. Similar to the previous works, we selected two commonly used properties: penalized octanol-water partition coefficient (penalized logP) \cite{kusner2017grammar} and quantitative estimation of drug-likeness (QED) \cite{Gomez-Bombarelli2018-jp}. We considered genetic and predictor-guided optimization strategies.

For genetic optimization, we start by sampling $256$ random molecules from ZINC250k dataset and computing their latent codes. Then we hierarchically optimize latent codes for $3000$ iterations. At each iteration we generate a new population using crossing-over and mutation and keep $256$ molecules with the highest reward. In crossing-over, we randomly permute all molecules in the population to form $256$ pairs. For each pair, we uniformly sample latent codes from spherical linear interpolation (Slerp) trajectory \cite{white2016sampling} and reconstruct the resulting molecule. We mutate one level's latent code at each iteration. Starting with a top level, we resample $10\%$ of the components from a Gaussian distribution. For genetic optimization, we compare different mutation and crossing-over strategies, including top level optimization with fixed bottom layers and vice versa.

In predictor-guided optimization, we followed the approach proposed by \citet{Jin2018-sm}: we fine-tuned the pre-trained model jointly with a penalized logP predictor from the high-level latent codes for one epoch (MAE=$0.41$ for penalized logP, MAE=$0.07$ for QED). We randomly sampled $2560$ molecules from a prior distribution and took $200$ constrained gradient ascent steps along the predictor's gradient to modify the high-level latent codes; we resample low-level latent codes from the prior. We decrease the learning rate after each iteration and keep the best reconstructed molecule that falls into the constrained region. Intuitively, the gradient ascent over high-level latent codes guides the search towards better global structure, while low-level latent codes produce a diverse set of molecules with the same global structure and similar predicted values.

We report the scores of the best molecules found during optimization in Table~\ref{tab:global} and provide optimization trajectories in supplementary information C. 

\begin{table*}[h]
\caption{Molecular property optimization: penalized octanol-water partition coefficient (penalized logP) and quantitative estimation of drug-likeness (QED). Results for baseline models from \cite{shi2020graphaf, madhawa2019graphnvp}.}
\label{tab:global}
\centering
\resizebox{\textwidth}{!}{
\begin{tabular}{lllllll}
\toprule

\multirow{2}{*}{Method} &
\multicolumn{3}{c}{Penalized logP} &
\multicolumn{3}{c}{QED} \\

\cmidrule(lr){2-4} \cmidrule(l){5-7}
 & 1st & 2nd & 3rd  & 1st & 2nd & 3rd \\ 
\midrule
ZINC250k & 4.52 & 4.30 & 4.23 & \textbf{0.948} & \textbf{0.948} & \textbf{0.948} \\
\midrule
\multicolumn{7}{c}{Graph-based models} \\ 
\midrule
GCPN \cite{you2018graph} & 7.98 & 7.85 & 7.80 & \textbf{0.948} & 0.947 & 0.946 \\
MolecularRNN \cite{popova2019molecularrnn} & 8.63 & 6.08 & 4.73 & 0.844 & 0.796 & 0.736 \\
GraphNVP \cite{madhawa2019graphnvp} & - & - & - & 0.833 & 0.723 & 0.706 \\
GraphAF \cite{shi2020graphaf} & 12.23 & 11.29 & 11.05 & \textbf{0.948} & \textbf{0.948} & \textbf{0.948} \\
MoFlow \cite{zang2020moflow} & - & - & - & \textbf{0.948}  & \textbf{0.948}  & \textbf{0.948}  \\
\midrule
\multicolumn{7}{c}{Proposed model} \\
\midrule
MolGrow (GE) & $\mathbf{14.01 \pm 0.364}$ & $\mathbf{13.95 \pm 0.424}$ & $\mathbf{13.92 \pm 0.422}$ & $\mathbf{0.9484 \pm 0.0}$ & $\mathbf{0.9484 \pm 0.0}$ & $\mathbf{0.9484 \pm 0.0}$ \\
MolGrow (GE, Top only) & $11.66 \pm 0.31$ & $11.65 \pm 0.319$ & $11.63 \pm 0.306$ & $\mathbf{0.9484 \pm 0.0}$ & $\mathbf{0.9484 \pm 0.0}$ & $\mathbf{0.9484 \pm 0.0}$  \\
MolGrow (GE, Bottom only) & $10.29 \pm 3.32$ & $10.29 \pm 3.33$ & $10.28 \pm 3.32$ &$ \mathbf{0.9484 \pm 0.0}$ & $\mathbf{0.9484 \pm 0.0}$ & $\mathbf{0.9484 \pm 0.0}$  \\
MolGrow (predictor-guided optimization) & $5.2 \pm 0.347$ & $4.94 \pm 0.262$ & $4.84 \pm 0.22$ & $\mathbf{0.9484 \pm 0.0}$ & $0.9483 \pm 0.0$ & $0.9483 \pm 0.0$ \\
MolGrow (REINFORCE) & $4.81 \pm 0.285$ & $4.47 \pm 0.145$ & $4.39 \pm 0.126$  & $0.9468 \pm 0.001$ & $0.9459 \pm 0.001$ & $0.9455 \pm 0.001$ \\
\midrule
\multicolumn{7}{c}{SMILES and fragment-based models} \\
\midrule
DD-VAE \cite{pmlr-v108-polykovskiy20a} & 5.86 & 5.77 & 5.64 & - & - & - \\
Grammar VAE \cite{kusner2017grammar} & 2.94 & 2.88 & 2.80 & - & - & - \\
SD-VAE \cite{dai2018syntax} & 4.04 & 3.50 & 2.96 & - & - & - \\
JT-VAE \cite{Jin2018-sm} & 5.30 & 4.93 & 4.49 & \textbf{0.948} & 0.947 & 0.947 \\
\bottomrule
\end{tabular}
}
\end{table*}

\subsection{Constrained optimization}
In this section, we apply MolGrow to constrained molecular optimization. In this task, we optimize a chemical property in proximity of the initial molecule. Following the previous works \cite{Jin2018-sm, you2018graph}, we selected $800$ molecules with the lowest penalized octanol-water partition coefficient (logP) and constrain minimum Tanimoto similarity $\delta$ between Morgan fingerprints \cite{rogers2010extended} of the initial and final molecules. For constrained optimization, we followed the predictor-guided approach described above and optimize each of $800$ starting molecules for $200$ steps. In Table~\ref{tab:constrained}, we report average penalized logP improvement and similarity to the initial molecule. We also report a fraction of molecules for which we successfully discovered a new molecule with higher penalized logP. Note that unlike GCPN and GraphAF baselines, we do not fine-tune the model for each starting molecule, reducing time and memory costs for optimization.

\begin{table}[h]
\caption{Constrained optimization of penalized octanol-water partition coefficient (logP). Mean$\pm$std over $800$ initial molecules with the worst penalized logP in ZINC250k.}
\label{tab:constrained}
\centering
\resizebox{\columnwidth}{!}{
\begin{tabular}{cllllll}
\toprule
\multirow{2}{*}{$\delta$}  & \multicolumn{3}{c}{GCPN} 
& \multicolumn{3}{c}{GraphAF}                            \\ 
\cmidrule(r){2-4}  \cmidrule(lr){5-7}
& Improvement & Similarity & Success & Improvement & Similarity & Success \\
\midrule
0.0                       & $4.20 \pm 1.28$ & $0.32 \pm 0.12$ & 100\%  & $13.13 \pm 6.89$ & $0.29 \pm 0.15$ & 100\%  \\
0.2                       & $4.12 \pm 1.19$ & $0.32 \pm 0.11$ & 100\%  & $11.90 \pm 6.86$ & $0.33 \pm 0.12$ & 100\% \\
0.4                        & $2.49 \pm 1.30$ & $0.47 \pm 0.08$ & 100\%  & $8.21 \pm 6.51$  & $0.49 \pm 0.09$ & 99.88\% \\
0.6                        & $0.79 \pm 0.63$ & $0.68 \pm 0.08$ & 100\% & ${\bf4.98 \pm 6.49}$  & $0.66 \pm 0.05$ & 96.88\% \\ \bottomrule
\end{tabular}
}
\resizebox{\columnwidth}{!}{
\begin{tabular}{cllllll}
\toprule
\multirow{2}{*}{$\delta$}  &   \multicolumn{3}{c}{MoFlow} & \multicolumn{3}{c}{MolGrow}                      \\ \cmidrule(r){2-4}  \cmidrule(lr){5-7}
& Improvement & Similarity & Success & Improvement & Similarity & Success\\
\midrule
0.0                       & $8.61 \pm 5.44$ & $0.30 \pm 0.20$ & 98.88\%  & ${\bf 14.84 \pm 5.786}$ & $ 0.048 \pm 0.038$ & 100\%  \\
0.2                       & $7.06 \pm 5.04$ & $0.43 \pm 0.20$ & 96.75\%  & ${\bf 11.99 \pm 6.45}$ & $0.23 \pm 0.045$   & 99.88\%  \\
0.4                        & $4.71 \pm 4.55$ & $0.61 \pm 0.18$ & 85.75\% & ${\bf 8.337 \pm 6.85 }$ & $0.44 \pm 0.048$ & 99.88\%  \\
0.6                        & $2.10 \pm 2.86$ & $0.79 \pm 0.14$ & 58.25\% & $ 4.063 \pm 5.609 $ & $0.65 \pm 0.068$ & 97.78\%   \\ \bottomrule
\end{tabular}
}
\end{table}

\section{Conclusion}
In this paper, we presented a new hierarchical molecular graph generative model and outperformed existing node-level models on distribution learning and molecular property optimization tasks. On distribution learning, string- and fragment-based generators still perform better than node-level models, since they explicitly handle valency and connectivity constraints. Similar to the previous models, we obtained better performance when learning on a fixed atom ordering. Our fragment-oriented ordering further improves the results over BFS. In this work, we compared generated and test sets using standard distribution learning metrics and found out that the distributions produced by previous node-level graph generators differ significantly from the test set, although these models were trained for distribution learning.

\clearpage

\appendix

\section{Appendix A. Hyperparameters}
MolGrow model has $6$ steps for ZINC250K dataset,  $5$ steps for MOSES dataset, and $4$ steps for QM9 dataset. Each step contains $2$ blocks if its input graph contains at most $16$ nodes, and $4$ blocks otherwise. We apply additional $4$ blocks to the initial one-node latent codes.

CAGE projects node and edge feature vectors onto a $64$ dimensional manifold; the number of attention heads equals $16$. Neural networks $s^v_{\theta}$, $t^v_{\theta}$, $s^e_{\theta}$, and $t^e_{\theta}$ are $2$-layer fully-connected neural networks with ReLU activations. The hidden size of these networks is $4$ times bigger than the output size. The final layer of $s^e_{\theta}$ and $s^v_{\theta}$ is a sigmoid function.

In $QR$ decomposition for linear layers we parameterize matrix $Q$ decomposition with $128$ Householder reflections. We sample $v_i$ from $\mathcal{N}(0, I)$ and normalize them. We re-normalize $v_i$ after each model optimization step. We initialize matrix $R$ with elements from $\mathcal{N}(0, 0.05 * I)$ above diagonal; zeros below the diagonal, and ones on diagonal.

We train the model with Adam optimizer with learning rate $0.001$ which decreases by a factor of $0.8$ after each training epoch for MOSES, and by a factor of $0.92$ after each training epoch for QM9 and ZINC250K datasets. The batch size is $256$ for MOSES and QM9 datasets, and $80$ for ZINC250K datasets. We train the model for $10$ epochs on MOSES and $30$ epochs on ZINC250k and QM9.

In our experiments we used QM9, ZINC250k, and MOSES datasets. For QM9 we use $N=16$ nodes with $d_v=6$ atom types, including padding atom type. For MOSES we used $N=32$ nodes with $d_v=9$ atom types, and $N=64$ with $d_v=14$ ($11$ for atom types and $3$ for charge type) and for ZINC250K. The number of bond types is $d_e=4$ (none, single, double, tripple).

We sample the latent codes from a multivariate normal distribution $\mathcal{N}(0, I)$ and multiply them by a temperature parameter $0.7$, similar to the previous works. 

We performed a hyperparameter search on the number of blocks in each level (considered $2, 4, 8, 16$ blocks) and internal shape of CAGE (considered $8, 16, 32, 64$ elements in hidden layer). The best hyperparameters are described above. To train MolGrow on one dataset we used Tesla K80 GPU. The training procedure took approximately 2 days for each dataset.

\section{Appendix B. Distribution learning}
\begin{figure}[h]
\begin{center}
    \centerline{\includegraphics[width=0.9\columnwidth]{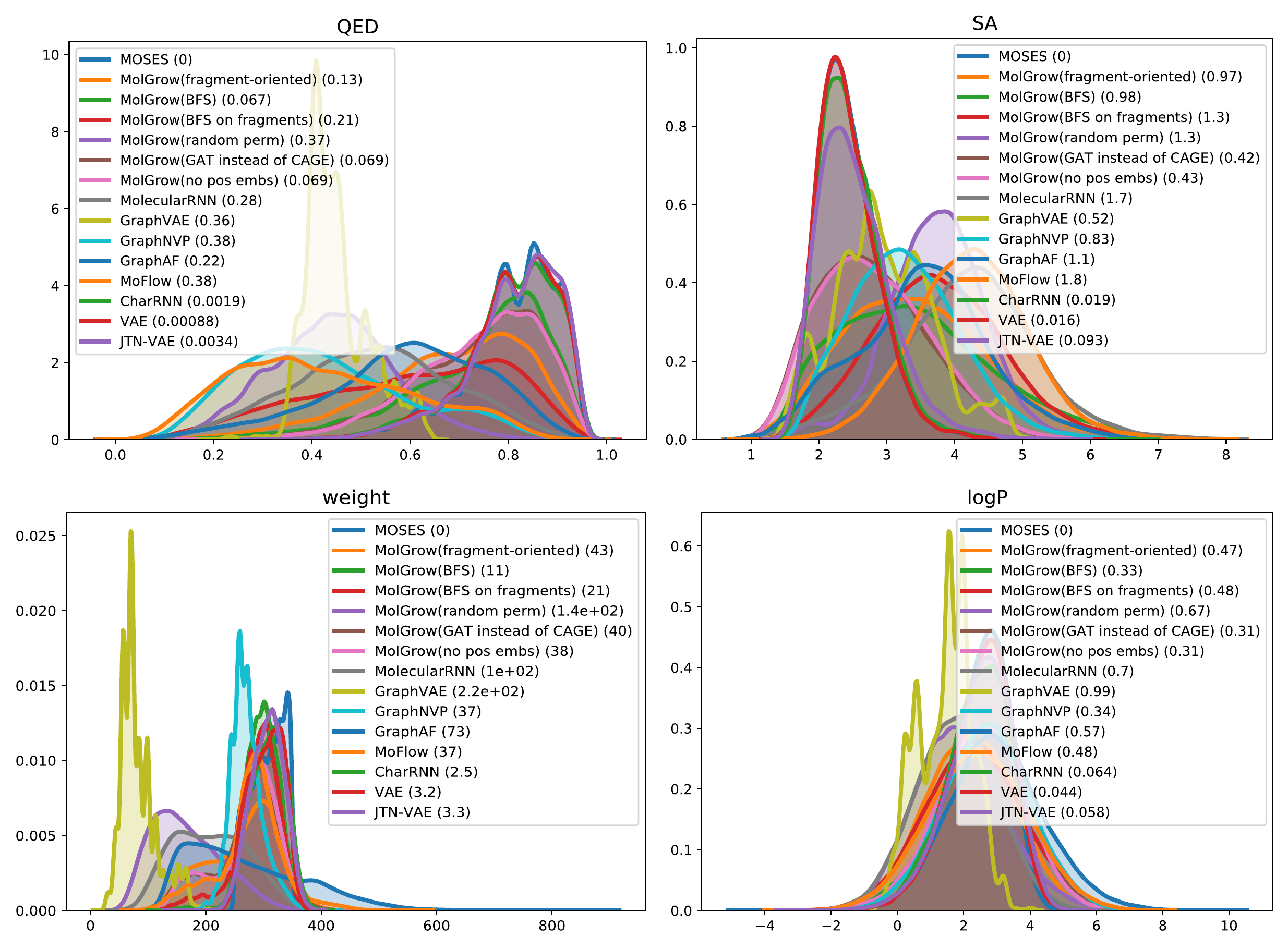}}
    \caption{Properties distribution for models trained on MOSES. Quantitative estimate of drug-likeness (QED), synthetic accessibility score (SA), molecular weight (weight), and octanol-water partition coefficient (logP).}
    \label{fig:moses_distributions}
\end{center}
\end{figure}

In this section, we provide additional experiments on QM9 \cite{ramakrishnan2014quantum} and ZINC250k \cite{kusner2017grammar} datasets (Table~\ref{tab:zinc} and Table~\ref{tab:qm9}). To compute validity for MolGrow, we removed edges that exceed maximum atom valency. If a final graph contains more than one fragment, we keep only the biggest one. Validity without cleanup corresponds to a fraction of valid molecules when no additional manipulations are performed on the generated graph. We check validity using RDKit and additionally consider molecules with multiple components invalid. We provide property distribution plots for models trained on MOSES in Figure~\ref{fig:moses_distributions}.

\section{Appendix C. Molecular property optimization}
In this section, we illustrate genetic optimization of molecular properties. In Figures \ref{fig:trajectory_logp} and \ref{fig:trajectory_qed}, we show best molecules in intermediate populations without consecutive duplicates. In Figures \ref{fig:best_logp} and \ref{fig:best_qed}, we show the best molecules in the final population.

\begin{figure*}[h]
\begin{center}
    \centerline{\includegraphics[width=1.9\columnwidth]{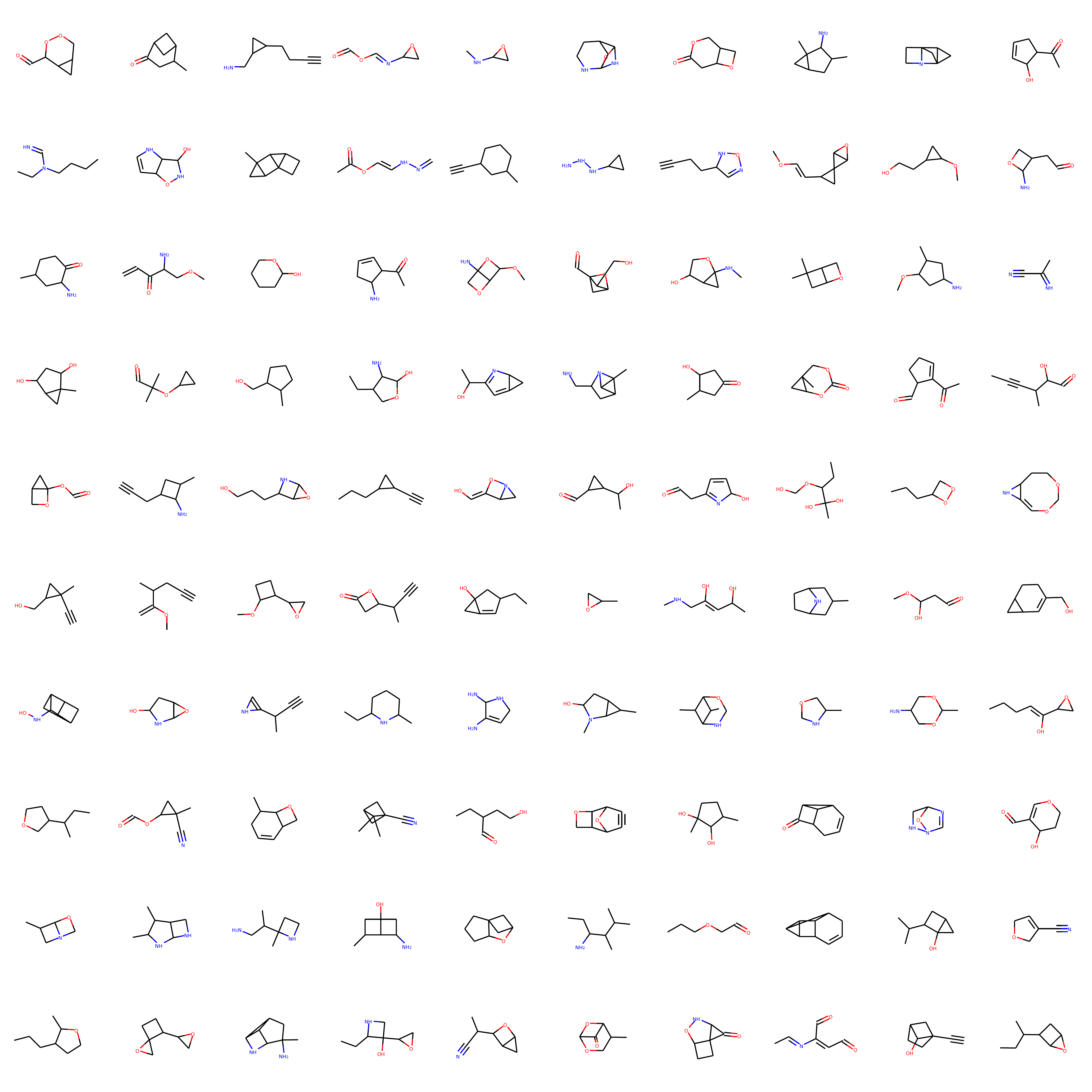}}
    \caption{Samples from MolGrow model trained on QM9 dataset.}
    \label{fig:sample_qm9}
\end{center}
\end{figure*}

\begin{figure*}[h]
\begin{center}
    \centerline{\includegraphics[width=1.8\columnwidth]{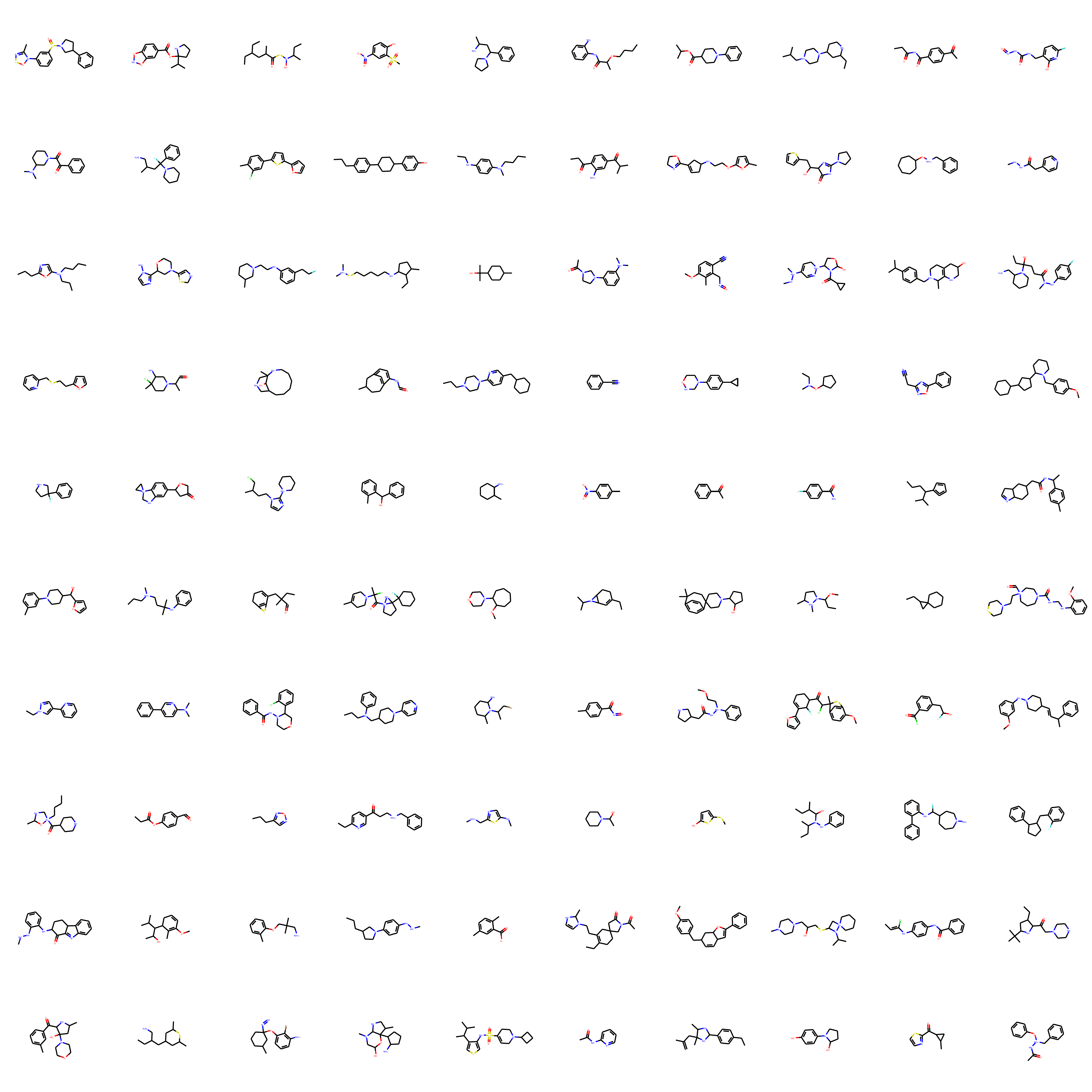}}
    \caption{Samples from MolGrow model trained on ZINC250K dataset.}
    \label{fig:sample_zinc}
\end{center}
\end{figure*}

\begin{figure*}[h]
\begin{center}
    \centerline{\includegraphics[width=1.8\columnwidth]{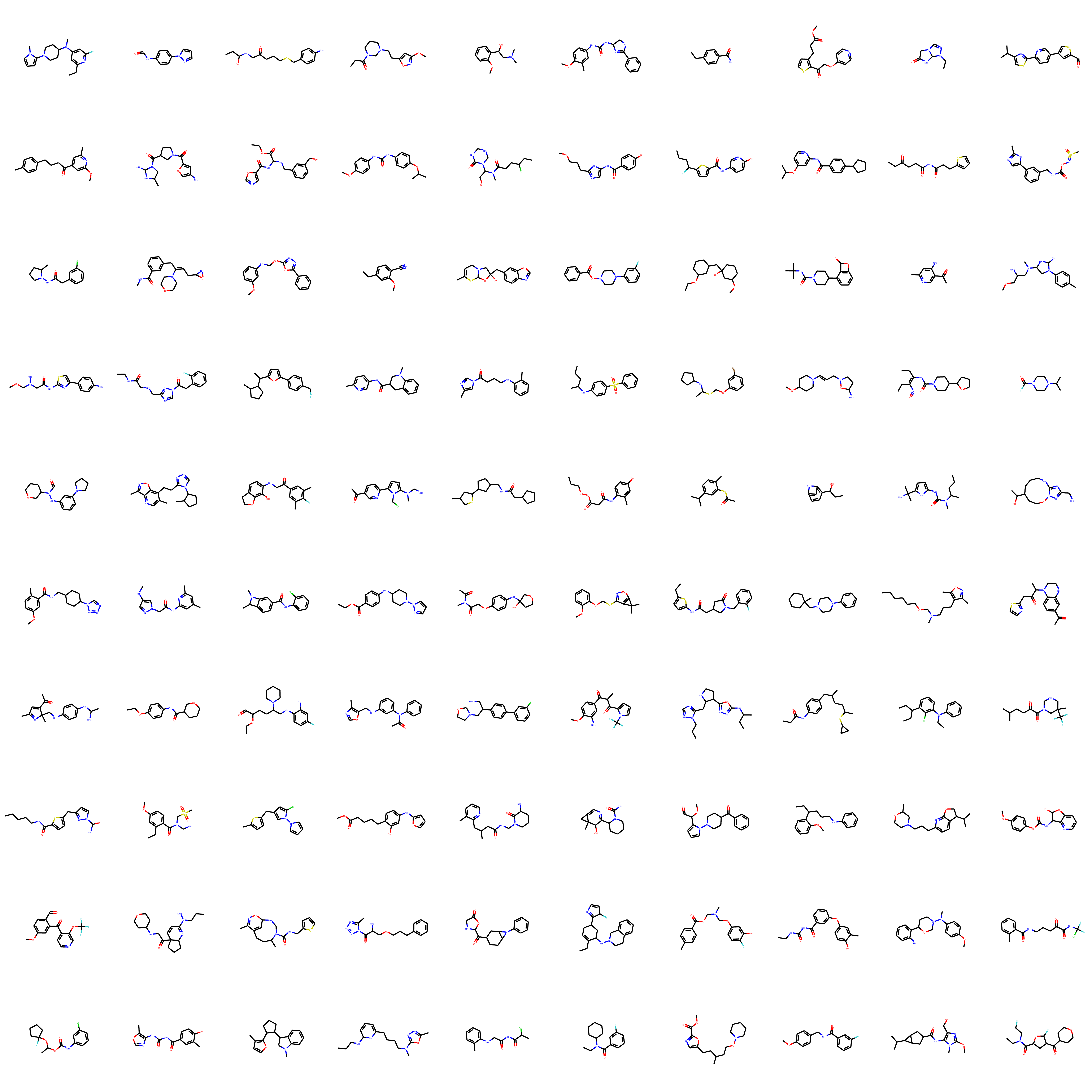}}
    \caption{Samples from MolGrow model trained on MOSES dataset.}
    \label{fig:sample_moses}
\end{center}
\end{figure*}

In Figure~\ref{fig:sample_qm9}, Figure~\ref{fig:sample_zinc}, and Figure~\ref{fig:sample_moses} we provide random samples from MolGrow trained on QM9, ZINC and MOSES datasets correspondingly.

\begin{table*}[ht]
\caption{Results for distribution learning on ZINC250k dataset. Baseline results from \cite{Jin2018-sm, popova2019molecularrnn, madhawa2019graphnvp, shi2020graphaf, zang2020moflow}}
\label{tab:zinc}
\centering
\begin{tabular}{llllll}
\hline
Method       & Validity & Validity without cleanup & Unique@10k & Novelty & Reconstruction \\ \hline
JT-VAE       & 100\%    & -                       & 100\%      & 100\%   & 76.7\%         \\
MolecularRNN & 100\%    & 65\%                    & 99.89\%    & 100\%   & -              \\
GraphNVP     & 42.6\%   & -                       & 94.8\%     & 100\%   & 100\%          \\
GraphAF      & 100\%    & 68\%                    & 99.1\%     & 100\%   & 100\%          \\
MoFlow  & $100 \pm 0.0 $\%    & $81.76 \pm 0.21$\%                    & $99.99 \pm 0.01$\%     & $100 \pm 0.0 $\%   & $100 \pm 0.0 $\% \\
MolGrow         & $100 \pm 0.0 $\%        & $57.8 \pm 7.75$\%                    & $99.06 \pm 0.46$\%          & $99.96 \pm 0.01$\%       & $100 \pm 0.0 $\%         \\ \hline
\end{tabular}
\end{table*}

\begin{table*}[ht]
\caption{Results for distribution learning on QM9 dataset. Baseline results from \cite{shi2020graphaf}.}
\label{tab:qm9}
\centering
\begin{tabular}{llllll}
\hline
Method   & Validity & Validity without cleanup & Unique@10k & Novelty & Reconstruction \\ \hline
GraphVAE  & -        & 55\%                    & 76\%       & 61\%    & 55\%           \\
GraphNVP & 83\%     & -                       & 99.2\%     & 58.2\%   & 100\%          \\
GraphAF  & 100\%    & 67\%                    & 94.5\%     & 88.8\%  & 100\%          \\
MoFlow  & $100 \pm 0.0 $\%    & $96.17 \pm 0.18 $\%                    & $99.20 \pm 0.12 $\%     & $98.03 \pm 0.14 $\% & $100 \pm 0.0 $\%          \\
MolGrow     & $100 \pm 0.0 $\%        & $86.80 \pm 0.12 $\%                    & $99.15 \pm 0.05 $\%          & $81.43 \pm 0.53 $\%      & $100 \pm 0.0 $\%          \\ \hline
\end{tabular}
\end{table*}

\begin{table*}[ht]
\caption{Performance metrics for baseline and proposed models on MOSES benchmark}
\label{tab:moses_metrics}
\centering
\resizebox{1.95 \columnwidth}{!}{
\begin{tabular}{llllllllllllllll}
\toprule
\multirow{2}{*}{Model}  & \multirow{2}{*}{Valid ($\uparrow$)}  & \multirow{2}{*}{Unique@1k ($\uparrow$)}  & \multirow{2}{*}{Unique@10k ($\uparrow$)}  & \multicolumn{2}{c}{FCD ($\downarrow$)} & \multicolumn{2}{c}{SNN ($\uparrow$)} & \multicolumn{2}{c}{Frag ($\uparrow$)} & \multicolumn{2}{c}{Scaf ($\uparrow$)} & \multirow{2}{*}{IntDiv ($\uparrow$)}  & \multirow{2}{*}{IntDiv2 ($\uparrow$)}  & \multirow{2}{*}{Filters ($\uparrow$)}  & \multirow{2}{*}{Novelty ($\uparrow$)} \\
 &  &  &  & Test & TestSF & Test & TestSF & Test & TestSF & Test & TestSF &  &  &  & \\
\midrule
\multicolumn{15}{c}{Graph-based models} \\ 
\midrule
                 MolecularRNN &  {\bf 1.0 } &            0.999 &           0.9861 &             23.1302 &             24.1749 &              0.279 &              0.2681 &              0.5657 &              0.5579 &              0.0457 &             0.0075 &  {\bf 0.9062 } &  {\bf 0.8985 } &              0.6143 &           0.9996 \\
                     GraphVAE &  {\bf 1.0 } &            0.207 &           0.0495 &             49.3917 &             50.7191 &             0.2082 &               0.189 &              0.2304 &              0.2148 &                 0.0 &                0.0 &              0.8238 &              0.7769 &               0.963 &  {\bf 1.0 } \\
                     GraphNVP &  {\bf 1.0 } &            0.999 &           0.9974 &             29.9517 &             31.2123 &             0.3057 &              0.2836 &              0.6297 &              0.6144 &              0.0266 &             0.0139 &              0.8574 &              0.8448 &              0.6008 &           0.9999 \\
                      GraphAF &  {\bf 1.0 } &            0.989 &           0.9707 &               21.85 &             23.2515 &             0.3056 &              0.2891 &              0.6513 &              0.6346 &              0.1255 &             0.0137 &              0.9036 &              0.8927 &              0.4145 &           0.9999 \\
                      MoFlow &  {\bf 1.0 } &        {\bf 1.0 } &          {\bf 1.0 } &               28.05 &             29.30 &             0.2554 &              0.2426 &              0.6858 &              0.6683 &              0.0224 &             0.0068 &              0.8923 &              0.8849 &              0.4126 &           0.9999 \\
\midrule
\multicolumn{15}{c}{Proposed model} \\ 
\midrule
   MolGrow(fragment-oriented) &  {\bf 1.0 } &  {\bf 1.0 } &            0.998 &              5.7888 &              6.3861 &             0.4335 &              0.4208 &              0.9518 &              0.9464 &              0.6697 &             0.0769 &              0.8663 &              0.8599 &              0.8274 &           0.9938 \\
                 MolGrow(BFS) &  {\bf 1.0 } &  {\bf 1.0 } &           0.9999 &              9.2024 &              9.8472 &             0.3997 &              0.3891 &              0.9476 &              0.9414 &              0.6628 &             0.0631 &              0.8724 &              0.8665 &              0.6673 &           0.9928 \\
    MolGrow(BFS on fragments) &  {\bf 1.0 } &  {\bf 1.0 } &  {\bf 1.0 } &             15.0707 &             15.9881 &             0.3379 &              0.3269 &              0.8902 &              0.8794 &              0.1875 &             0.0308 &              0.8883 &              0.8819 &              0.6169 &           0.9993 \\
         MolGrow(random perm) &  {\bf 1.0 } &            0.967 &           0.8861 &             36.6702 &             37.9284 &             0.2292 &              0.2136 &              0.0729 &              0.0696 &              0.0002 &                0.0 &              0.8735 &              0.8616 &              0.5777 &  {\bf 1.0 } \\
 MolGrow(GAT instead of CAGE) &  {\bf 1.0 } &  {\bf 1.0 } &           0.9953 &              6.4564 &              6.9704 &             0.4383 &              0.4263 &              0.9499 &              0.9449 &              0.6599 &             0.0827 &              0.8704 &               0.863 &              0.8367 &           0.9936 \\
 MolGrow(No positional embeddings) &  {\bf 1.0 } &  {\bf 1.0 } &           0.9926 &              7.5529 &              8.1794 &             0.4242 &              0.4138 &              0.9320 &              0.9279 &              0.6029 &             0.0828 &              0.8744 &               0.8668 &              0.8177 &           0.9949 \\
 \midrule
\multicolumn{15}{c}{SMILES and fragment-based models} \\ 
\midrule
                      CharRNN &           0.9884 &  {\bf 1.0 } &           0.9992 &  {\bf 0.0615 } &  {\bf 0.4967 } &             0.6109 &              0.5709 &  {\bf 0.9999 } &              0.9984 &              0.9269 &  {\bf 0.107 } &              0.8566 &              0.8507 &               0.996 &           0.8232 \\
                          VAE &           0.9766 &  {\bf 1.0 } &           0.9982 &              0.1094 &              0.5548 &  {\bf 0.626 } &  {\bf 0.5776 } &              0.9993 &  {\bf 0.9987 } &  {\bf 0.9411 } &             0.0485 &              0.8556 &              0.8496 &  {\bf 0.9971 } &           0.6871 \\
                      JTN-VAE &  {\bf 1.0 } &  {\bf 1.0 } &           0.9992 &              0.4224 &              0.9962 &             0.5561 &              0.5273 &              0.9962 &              0.9948 &              0.8925 &             0.1005 &              0.8512 &              0.8453 &              0.9778 &           0.9153 \\
\bottomrule
\end{tabular}
}
\end{table*}

\begin{figure*}[h]
\begin{center}
    \centerline{\includegraphics[width=1.8\columnwidth]{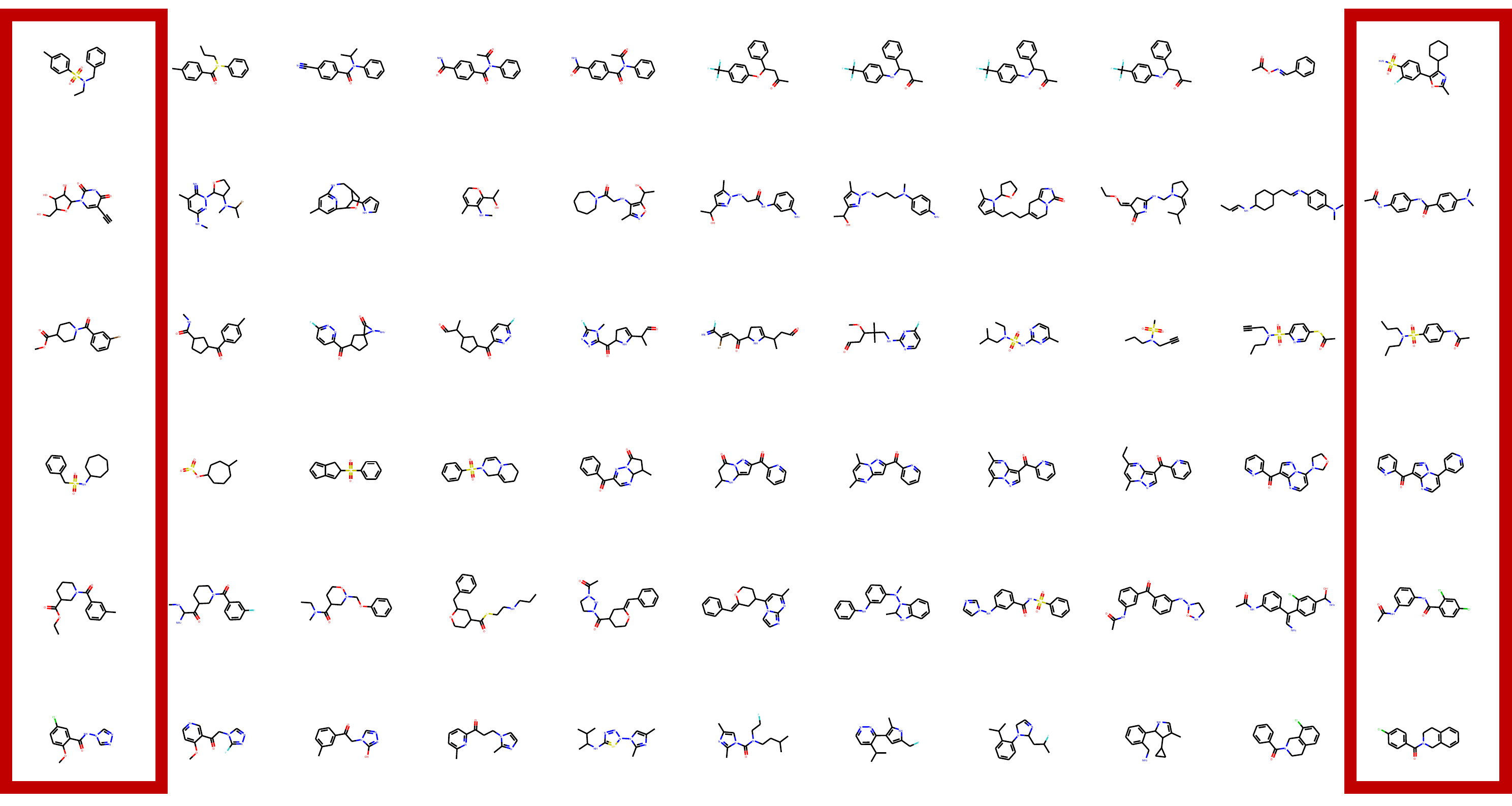}}
    \caption{Spherical interpolation between two molecules in the latent space of MolGrow model trained on MOSES dataset.}
    \label{fig:interpolation_moses}
\end{center}
\end{figure*}

\begin{figure*}[h]
\begin{center}
    \centerline{\includegraphics[width=1.4\columnwidth]{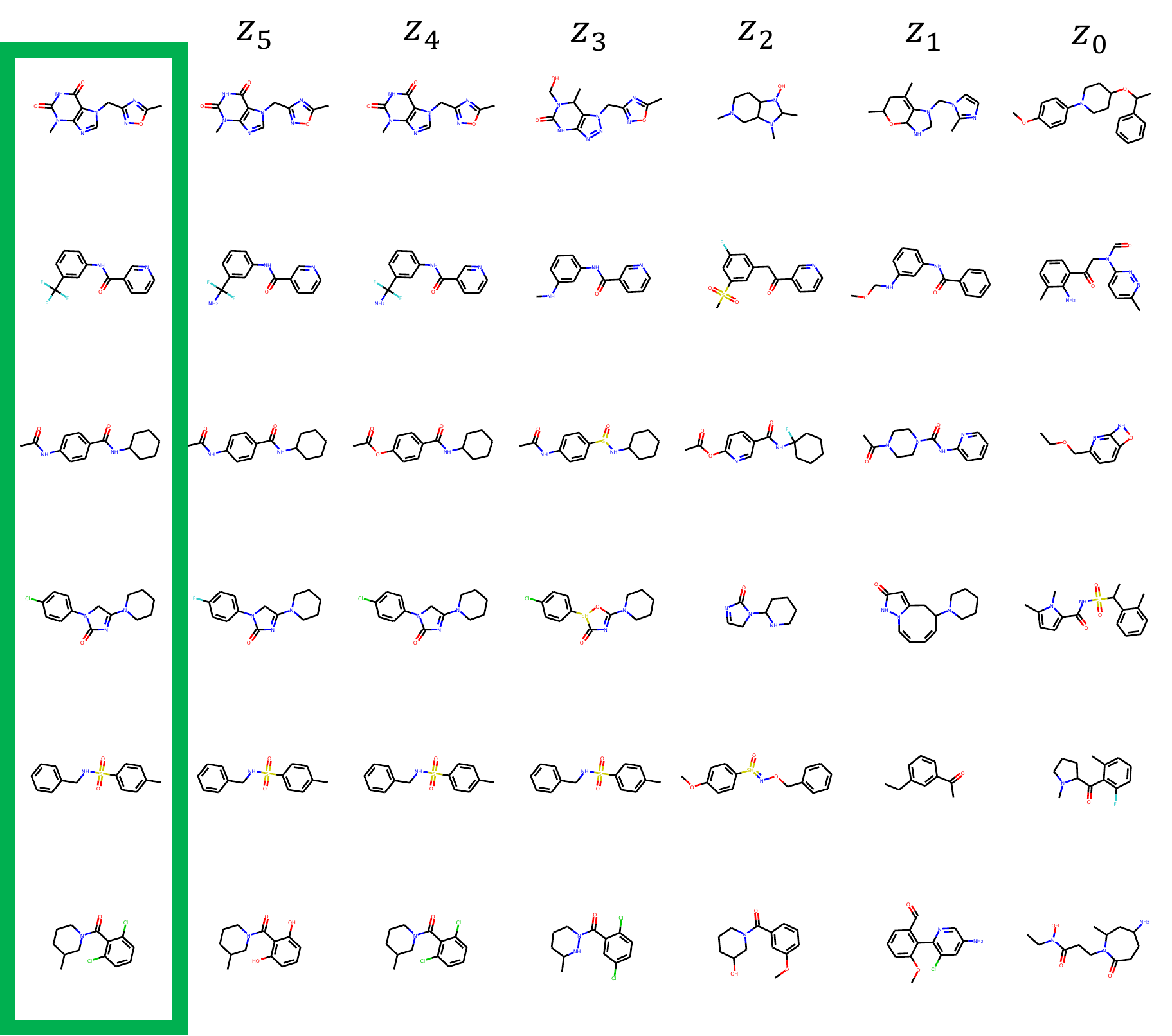}}
    \caption{Resampling of latent codes on different hierarchical levels of MolGrow model trained on MOSES dataset.}
    \label{fig:resampling_moses}
\end{center}
\end{figure*}

\begin{figure*}[h]
\begin{center}
    \centerline{\includegraphics[width=1.8\columnwidth]{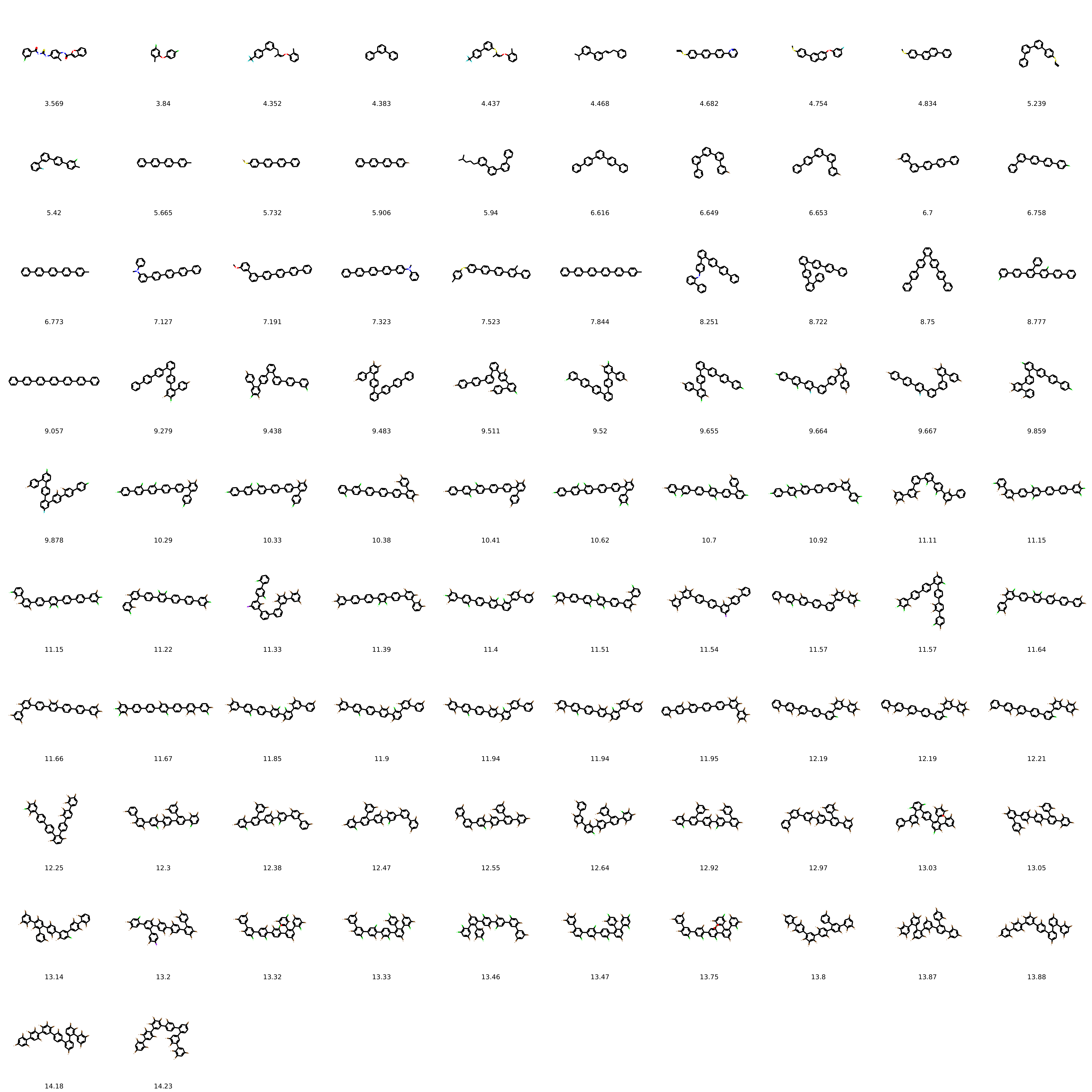}}
    \caption{Best molecules in intermediate populations for penalized octanol-water partition coefficient (logP).}
    \label{fig:trajectory_logp}
\end{center}
\end{figure*}

\begin{figure*}[h]
\begin{center}
    \centerline{\includegraphics[width=1.8\columnwidth]{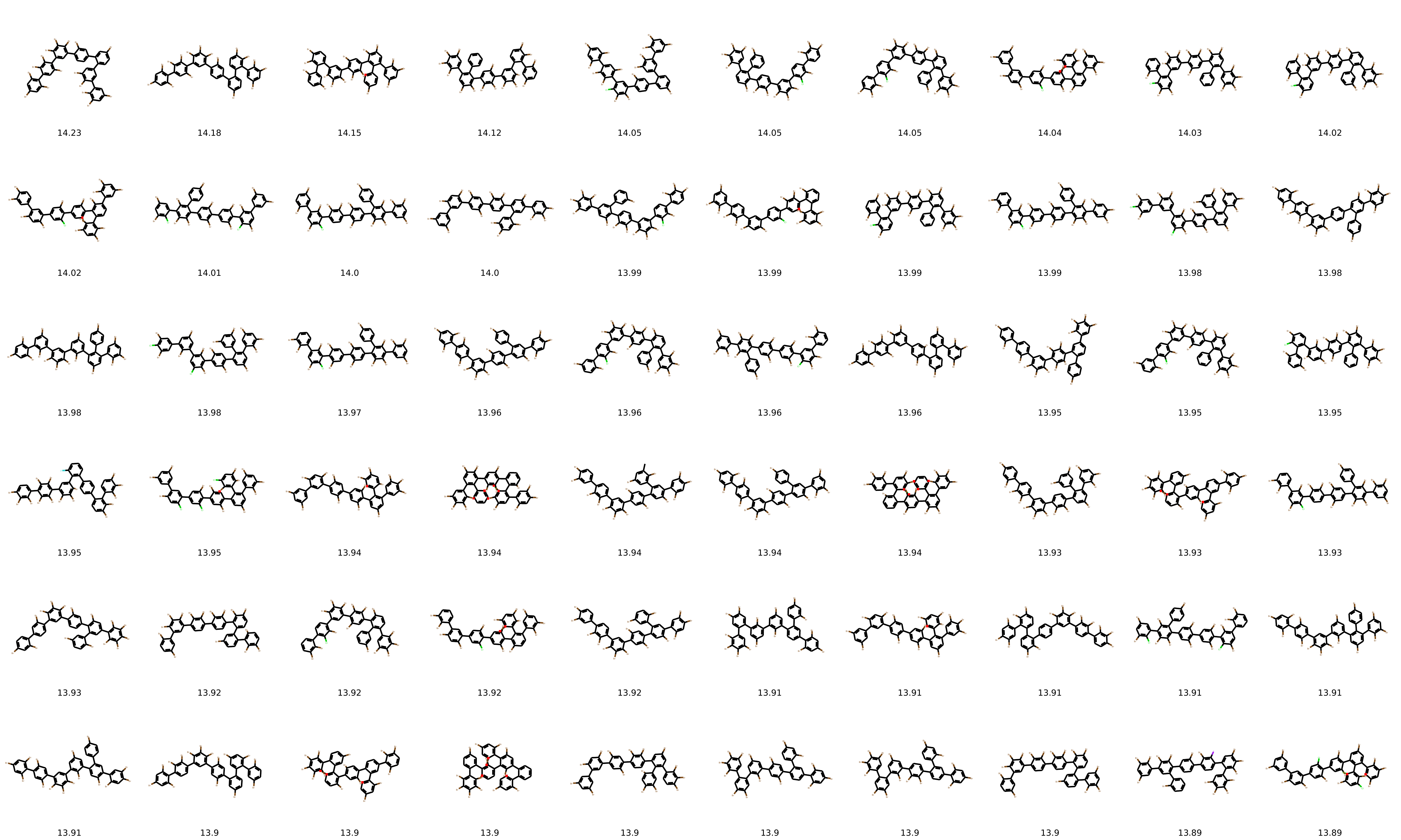}}
    \caption{Best molecules in the final population during genetic optimization of penalized octanol-water partition coefficient (logP).}
    \label{fig:best_logp}
\end{center}
\end{figure*}

\begin{figure*}[h]
\begin{center}
    \centerline{\includegraphics[width=1.8\columnwidth]{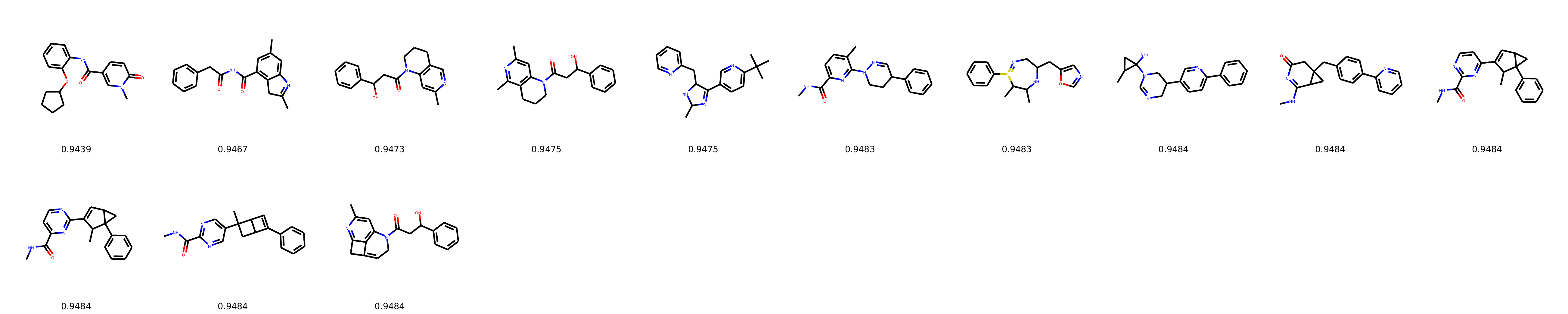}}
    \caption{Best molecules in intermediate populations for drug-likeness (QED).}
    \label{fig:trajectory_qed}
\end{center}
\end{figure*}

\begin{figure*}[h]
\begin{center}
    \centerline{\includegraphics[width=1.7\columnwidth]{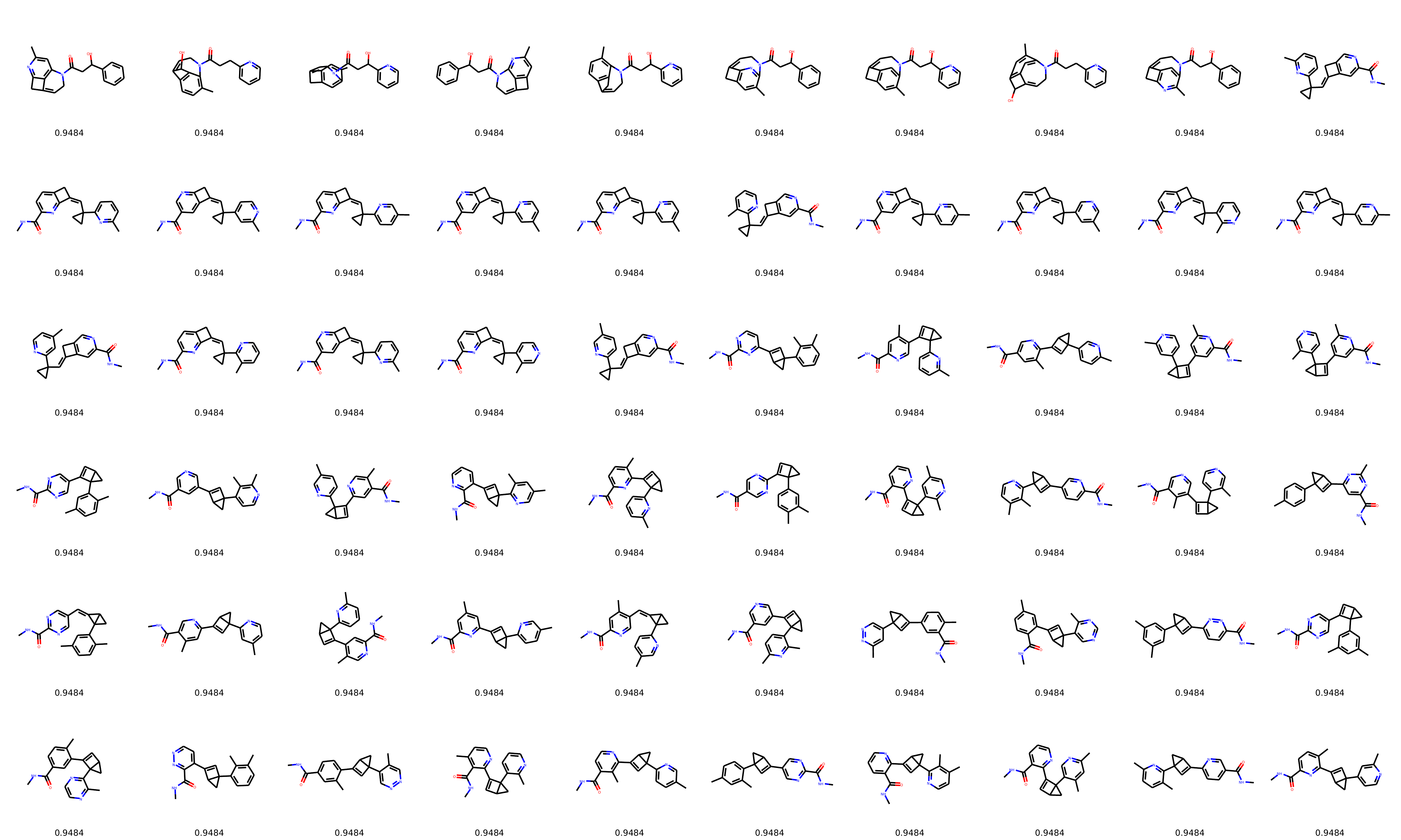}}
    \caption{Best molecules in the final population during genetic optimization of quantitative estimation of drug-likeness (QED).}
    \label{fig:best_qed}
\end{center}
\end{figure*}

 \begin{figure*}[h]
 \begin{center}
     \centerline{\includegraphics[width=1.7\columnwidth]{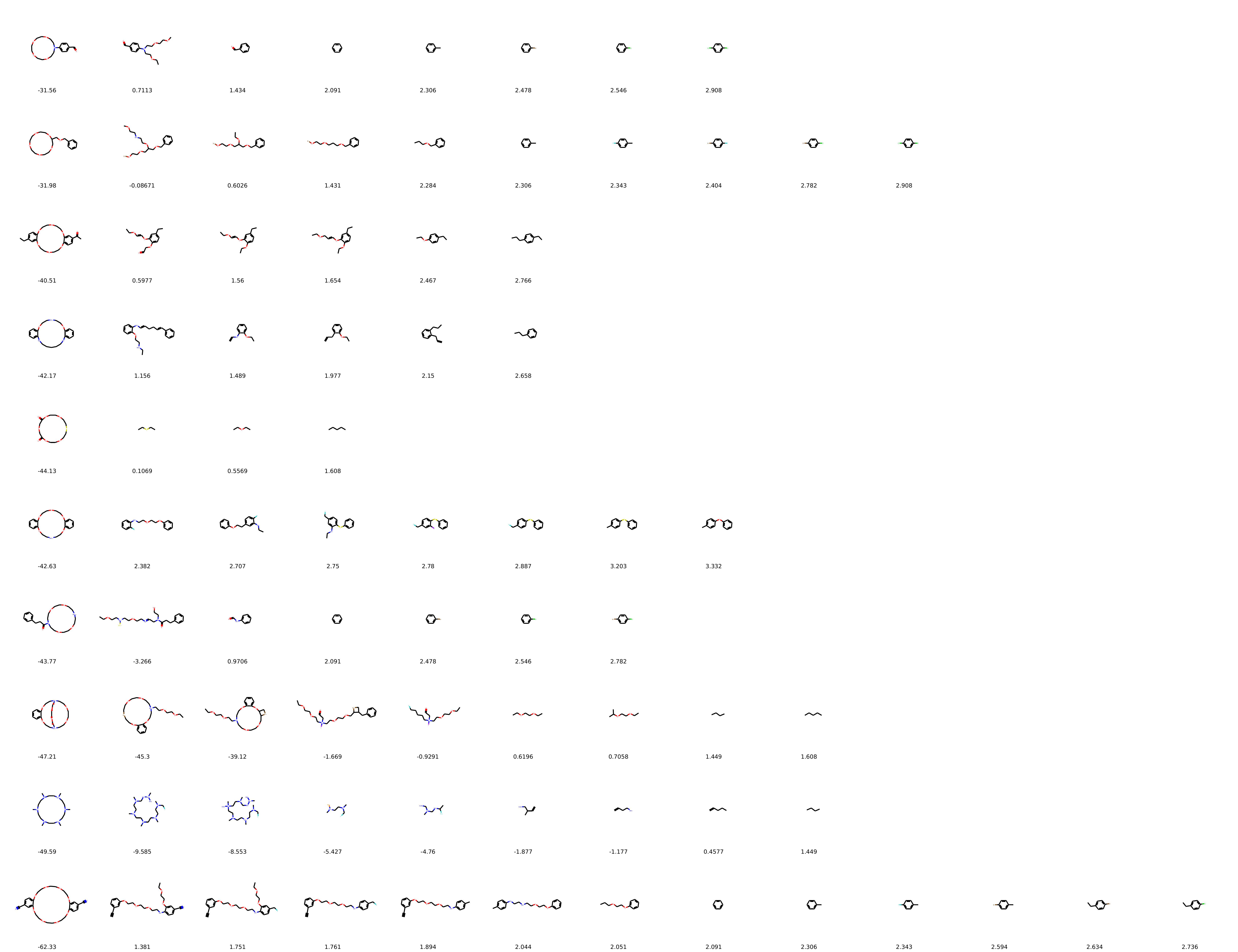}}
     \caption{Results of constrained optimization for  penalized octanol-water partition coefficient (logP): trajectories for molecules with best improvement of property.}
     \label{fig:constrained}
 \end{center}
 \end{figure*}

\end{document}